# Notes on Fermi-Dirac Integrals
## 4th Edition

Raseong Kim, Xufeng Wang, and Mark Lundstrom
Network for Computational Nanotechnology
Purdue University

**Preface to the 4th Edition:**

This edition is identical to the previous edition except for one change. The previous edition contained three Matlab scripts for computing Fermi-Dirac integrals [1]. The first two, *FD_int_approx.m* and *FDjx.m*, used analytic approximations and the third, *FD_int_num.m*, obtained Fermi-Dirac integrals by numerical integration. Because the third script was slow, the first two scripts were widely used, but errors can occur if the value of the input parameter, $\eta_F$, is outside of some bounds. This edition of the notes contains a new Matlab script, *TBFD_integral.m*, which obtains Fermi-Dirac integrals by interpolating from a pre-computed table. The table was generated using the GNU Scientific Library (GSL): https://www.gnu.org/software/gsl/doc/html/specfunc.html#fermi-dirac-function.

The use of the new routine is recommended for most users. It is nearly as accurate as the numerical integration routine and nearly as fast as the analytic approximation routine, but it is less likely to produce errors. (See Section 7 for benchmarking results.) The Matlab script can be downloaded at https://github.com/wang159/FDIntegral_Table.

## 1. Introduction

Fermi-Dirac integrals appear frequently in semiconductor problems, so a basic understanding of their properties is essential. The purpose of these notes is to collect in one place, some basic information about Fermi-Dirac integrals and their properties. We also present Matlab functions (see Appendix and [1]) that calculate Fermi-Dirac integrals (the "script F" defined by Dingle [2] and reviewed by Blakemore [3]) in three different ways.

To see how they arise, consider computing the equilibrium electron concentration per unit volume in a three-dimensional semiconductor with a parabolic conduction band from the expression,

$$n = \int_{E_C}^{\infty} g(E) f_0(E) dE = \int_{E_C}^{\infty} \frac{g(E)dE}{1 + e^{(E-E_F)/k_B T}}, \tag{1}$$

where $g(E)$ is the density of states, $f_0(E)$ is the Fermi function, and $E_C$ is the conduction band edge. For three dimensional electrons,

$$g_{3D}(E) = \frac{\left(2m^*\right)^{3/2}}{2\pi^2\hbar^3}\sqrt{E - E_C}\,, \tag{2}$$

which can be used in Eq. (1) to write

$$n = \frac{\left(2m^*\right)^{3/2}}{2\pi^2\hbar^3}\int_{E_C}^{\infty}\frac{\sqrt{E - E_C}\,dE}{1 + e^{(E-E_F)/k_BT}}\,. \tag{3}$$

By making the substitution,

$$\varepsilon = \left(E - E_C\right)/k_BT\,, \tag{4}$$

Eq. (3) becomes

$$n = \frac{\left(2m^*k_BT\right)^{3/2}}{2\pi^2\hbar^3}\int_0^{\infty}\frac{\varepsilon^{1/2}d\varepsilon}{1 + e^{\varepsilon-\eta_F}}\,, \tag{5}$$

where we have defined

$$\eta_F \equiv \left(E_F - E_C\right)/k_BT\,. \tag{6}$$

By collecting up parameters, we can express the electron concentration as

$$n_0 = N_{3D}\frac{2}{\sqrt{\pi}}F_{1/2}\left(\eta_F\right), \tag{7}$$

where

$$N_{3D} = 2\left(\frac{2\pi m^*k_BT}{h^2}\right)^{3/2} \tag{8}$$

is the so-called effective density-of-states and

$$F_{1/2}(\eta_F) \equiv \int_0^{\infty}\frac{\varepsilon^{1/2}d\varepsilon}{1 + \exp(\varepsilon - \eta_F)} \tag{9}$$

is the Fermi-Dirac integral of order 1/2. This integral can only be evaluated numerically. Note that its value depends on $\eta_F$, which measures the location of the Fermi level with respect to the conduction band edge. It is more convenient to define a related integral,

$$\mathcal{F}_{1/2}(\eta_F) \equiv \frac{2}{\sqrt{\pi}} \int_0^{\infty} \frac{\varepsilon^{1/2} d\varepsilon}{1+\exp(\varepsilon-\eta_F)}, \tag{10}$$

so that Eq. (7) can be written as

$$n = N_{3D} \mathcal{F}_{1/2}(\eta_F). \tag{11}$$

It is important to recognize whether you are dealing with the "Roman" Fermi-Dirac integral or the "script" Fermi-Dirac integral.

There are many kinds of Fermi-Dirac integrals. For example, in two dimensions, the density-of-states is

$$g_{2D}(E) = \frac{m^*}{\pi \hbar^2}, \tag{12}$$

and by following a procedure like that one we used in three dimensions, one can show that the electron density per unit area is

$$n_S = N_{2D} \mathcal{F}_0(\eta_F), \tag{13}$$

where

$$N_{2D} = \frac{m^* k_B T}{\pi \hbar^2}, \tag{14}$$

and

$$\mathcal{F}_0(\eta_F) = \int_0^{\infty} \frac{\varepsilon^0 d\varepsilon}{1+e^{\varepsilon-\eta_F}} = \ln\left(1+e^{\eta_F}\right) \tag{15}$$

is the Fermi-Dirac integral of order 0, which can be integrated analytically.

Finally, in one dimension, the density-of-states is

$$g_{1D}(E) = \frac{\sqrt{2m^*}}{\pi \hbar} \sqrt{\frac{1}{E-E_C}}, \tag{16}$$

and the equilibrium electron density per unit length is

$$n_L = N_{1D} \mathcal{F}_{-1/2}(\eta_F), \tag{17}$$

where

$$N_{1D} = \frac{1}{\hbar}\sqrt{\frac{2m^* k_B T}{\pi}}, \tag{18}$$

and

$$\mathcal{F}_{-1/2}(\eta_F) = \frac{1}{\sqrt{\pi}}\int_0^\infty \frac{\varepsilon^{-1/2} d\varepsilon}{1+e^{\varepsilon-\eta_F}} \tag{19}$$

is the Fermi-Dirac integral of order $-1/2$, which must be integrated numerically.

## 2. General Definition

In the previous section, we saw three examples of Fermi-Dirac integrals. More generally, we define

$$\mathcal{F}_j(\eta_F) \equiv \frac{1}{\Gamma(j+1)}\int_0^\infty \frac{\varepsilon^j d\varepsilon}{1+\exp(\varepsilon-\eta_F)}, \tag{20}$$

where $\Gamma$ is the gamma function. The $\Gamma$ function is just the factorial when its argument is a positive integer,

$$\Gamma(n) = (n-1)! \quad (\text{for } n \text{ a positive integer}). \tag{21a}$$

Also

$$\Gamma(1/2) = \sqrt{\pi}, \tag{21b}$$

and

$$\Gamma(p+1) = p\Gamma(p). \tag{21c}$$

As an example, let's evaluate $\mathcal{F}_{1/2}(\eta_F)$ from Eq. (20):

$$\mathcal{F}_{1/2}(\eta_F) \equiv \frac{1}{\Gamma(1/2+1)}\int_0^\infty \frac{\varepsilon^{1/2} d\varepsilon}{1+e^{\varepsilon-\eta_F}}, \tag{22a}$$

so we need to evaluate $\Gamma(3/2)$. Using Eqs. (21b) and (21c), we find,

$$\Gamma(3/2) = \Gamma(1/2+1) = \frac{1}{2}\Gamma(1/2) = \frac{\sqrt{\pi}}{2}, \tag{22b}$$

so $\mathcal{F}_{1/2}(\eta_F)$ is evaluated as

$$\mathcal{F}_{1/2}(\eta_F) \equiv \frac{2}{\sqrt{\pi}} \int_0^\infty \frac{\varepsilon^{1/2} d\varepsilon}{1+e^{\varepsilon-\eta_F}}, \tag{22c}$$

which agrees with Eq. (10). For more practice, use the general definition, Eq. (20) and Eqs. (21a-c) to show that the results for $\mathcal{F}_0(\eta_F)$ and $\mathcal{F}_{-1/2}(\eta_F)$ agree with Eqs. (15) and (19).

## 3. Derivatives of Fermi-Dirac Integrals

Fermi-Dirac integrals have the property that

$$\frac{d\mathcal{F}_j}{d\eta_F} = \mathcal{F}_{j-1}, \tag{23}$$

which often comes in useful. For example, we have an analytical expression for $\mathcal{F}_0(\eta_F)$, which means that we have an analytical expression for $\mathcal{F}_{-1}(\eta_F)$,

$$\mathcal{F}_{-1} = \frac{d\mathcal{F}_0}{d\eta_F} = \frac{1}{1+e^{-\eta_F}}. \tag{24}$$

Similarly, we can show that there is an analytic expression for any Fermi-Dirac integral of integer order, *j*, for $j \le -2$,

$$\mathcal{F}_j(\eta_F) = \frac{e^{\eta_F}}{\left(1+e^{\eta_F}\right)^{-j}} P_{-j-2}\left(e^{\eta_F}\right), \tag{25}$$

where $P_k$ is a polynomial of degree *k*, and the coefficients $p_{k,i}$ are generated from a recurrence relation [4] (note that the relation in Eq. (26c) is missing in p.222 of [4])

$$p_{k,0} = 1 \tag{26a}$$

$$p_{k,i} = (1+i)\, p_{k-1,i} - (k+1-i)\, p_{k-1,i-1} \qquad i = 1,\ldots,k-1 \tag{26b}$$

$$p_{k,k} = -p_{k-1,k-1}. \tag{26c}$$

For example, to evaluate $\mathscr{F}_{-4}(\eta_F) = e^{\eta_F} \big/ \left(1+e^{\eta_F}\right)^4 \times P_2\left(e^{\eta_F}\right)$, polynomial coefficients are generated from eqns. (26a-c) as [4]

$$\begin{aligned} &p_{0,0} = 1 \\ &p_{1,0} = 1 \qquad p_{1,1} = -p_{0,0} = -1 \\ &p_{2,0} = 1 \qquad p_{2,1} = 2p_{1,1} - 2p_{1,0} = -4 \qquad p_{2,2} = -p_{1,1} = 1 \end{aligned} \tag{27}$$

and we find

$$\mathscr{F}_{-4}(\eta_F) = \frac{e^{\eta_F}}{\left(1+e^{\eta_F}\right)^4} \sum_{i=0}^{2} p_{2,i} e^{i\eta_F} = \frac{e^{\eta_F}}{\left(1+e^{\eta_F}\right)^4} \left(1 - 4e^{\eta_F} + e^{2\eta_F}\right). \tag{28}$$

## 4. Asymptotic Expansions for Fermi-Dirac Integrals

It is useful to examine Fermi-Dirac integrals in the non-degenerate ($\eta_F << 0$) and degenerate ($\eta_F >> 0$) limits. For the non-degenerate limit, the result is particularly simple,

$$\mathscr{F}_J(\eta_F) \to e^{\eta_F}, \tag{29}$$

which means that for all orders, *j*, the Fermi-Dirac integral approaches the exponential in the non-degenerate limit. To examine Fermi-Dirac integrals in the degenerate limit, we consider the complete expansion for the Fermi-Dirac integral for $j > -1$ and $\eta_F > 0$ [2, 5, 6]

$$\mathscr{F}_J(\eta_F) = 2\eta_F^{j+1} \sum_{n=0}^{\infty} \frac{t_{2n}}{\Gamma(j+2-2n)\eta_F^{2n}} + \cos(\pi j) \sum_{n=1}^{\infty} \frac{(-1)^{n-1} e^{-n\eta_F}}{n^{j+1}}, \tag{30}$$

where $t_0 = 1/2$, $t_n = \sum_{\mu=1}^{\infty} (-1)^{\mu-1} \big/ \mu^n = \left(1-2^{1-n}\right)\zeta(n)$, and $\zeta(n)$ is the Riemann zeta function. The expressions for the Fermi-Dirac integrals in the degenerate limit ($\eta_F >> 0$) come from Eq. (30) as $\mathscr{F}_j(\eta_F) \to \eta_F^{j+1} \big/ \Gamma(j+2)$ [7]. Specific results for several Fermi-Dirac integrals are shown below.

$$\mathscr{F}_{-1/2}(\eta_F) \to \frac{2\eta_F^{1/2}}{\sqrt{\pi}} \tag{31a}$$

$$\mathscr{F}_{1/2}(\eta_F) \to \frac{4\eta_F^{3/2}}{3\sqrt{\pi}} \tag{31b}$$

$$\mathscr{F}_1(\eta_F) \to \frac{1}{2}\eta_F^2 \tag{31c}$$

$$\mathcal{F}_{3/2}(\eta_F) \to \frac{8\eta_F^{5/2}}{15\sqrt{\pi}} \quad (31d)$$

$$\mathcal{F}_{2}(\eta_F) \to \frac{1}{6}\eta_F^3 \quad (31e)$$

The complete expansion in Eq. (30) can be related to the well-known Sommerfeld expansion [8, 9]. First, note that the integrals to calculate carrier densities in Eqs. (1) and (3) are all of the form

$$\int_{-\infty}^{\infty} H(E) f_o(E) dE \,. \quad (32)$$

If $H(E)$ does not vary rapidly in the range of a few $k_B T$ about $E_F$, then we can write the Tayler expansion of $H(E)$ about $E_F$ as [9]

$$H(E) = \sum_{n=0}^{\infty} \frac{d^n}{dE^n} H(E)|_{E=E_F} \frac{(E-E_F)^n}{n!} \,. \quad (33)$$

Using this Taylor series expansion, the integral in Eq. (32) can be written as (see [9] for a detailed derivation)

$$\int_{-\infty}^{\infty} H(E) f_o(E) dE = \int_{-\infty}^{E_F} H(E) dE + \sum_{n=1}^{\infty} (k_B T)^{2n} a_n \frac{d^{2n-1}}{dE^{2n-1}} H(E)|_{E=E_F} \,, \quad (34)$$

where

$$a_n = 2\left(1 - \frac{1}{2^{2n}} + \frac{1}{3^{2n}} - \frac{1}{4^{2n}} + \cdots\right), \quad (35)$$

and it is noted that $a_n = 2t_{2n}$. Equation (34) is known as the Sommerfeld expansion [8, 9]. Typically, the first term in the sum in Eq. (34) is all that is needed, and the result is

$$\int_{-\infty}^{\infty} H(E) f_o(E) dE \simeq \int_{-\infty}^{E_F} H(E) dE + \frac{\pi}{6} (k_B T)^2 H'(E_F) \,. \quad (36)$$

If we scale E by $k_B T$ in Eq. (34), $\varepsilon$ , then Eq. (34) becomes

$$\int_{0}^{\infty} H(\varepsilon) f_0(\varepsilon) d\varepsilon = \int_{0}^{\eta_F} H(\varepsilon) d\varepsilon + \sum_{n=1}^{\infty} a_n \frac{d^{2n-1}}{d\varepsilon^{2n-1}} H(\varepsilon)\Big|_{\varepsilon=\eta_F} \,. \quad (37)$$

Then the Sommerfeld expansion for the Fermi-Dirac integral of order $j$ can be evaluated by letting $H(\varepsilon) = \varepsilon^j / \Gamma(j+1)$, and the result is

$$\mathcal{F}_J(\eta_F) = 2\eta_F^{j+1}\sum_{n=0}^{\infty}\frac{t_{2n}}{\Gamma\left(j+2-2n\right)\eta_F^{2n}}. \quad (38)$$

Equation (34) is the same as Eq. (30) except that the second term in Eq. (30) is omitted [3]. In the degenerate limit, however, the second term in Eq. (30) vanishes, so the eqns. (30) and (34) give the same results as eqns. (31a-e).

## 5. Approximate Expressions for Common Fermi-Dirac Integrals

Fermi-Dirac integrals can be quickly evaluated by tabulation [2, 7, 10, 11] or analytic approximation [12-14]. We briefly mention some of the analytic approximations and refer the reader to a Matlab function. Bednarczyk *et al.* [12] proposed a single analytic approximation that evaluates the Fermi-Dirac integral of order $j = 1/2$ with errors less than 0.4 % [3]. Aymerich-Humet *et al.* [13, 14] introduced an analytic approximation for a general $j$, and it gives an error of 1.2 % for $-1/2 < j < 1/2$ and 0.7 % for $1/2 < j < 5/2$, and the error increases with larger $j$. The Matlab fuction, "FD_int_approx.m [1]," calculates the Fermi-Dirac integral defined in Eq. (10) with orders $j \geq -1/2$ using these analytic approximations. The source code of this relatively short function is shown in the Appendix.

If a better accuracy is required and a longer CPU time is allowed, then the approximations proposed by Halen and Pulfrey [15, 16] may be used. In this model, several approximate expressions are introduced based on the series expansion in Eq. (30), and the error is less than $10^{-5}$ for $-1/2 \leq j \leq 7/2$ [14]. The Matlab function, "FDjx.m [1]," is the main function that calculates the Fermi-Dirac integrals using this model. This function includes tables of coefficients, so it is not simple enough to be shown in the Appendix, but it can be downloaded from [1].

There also have been discussions on the simple analytic calculation of the inverse Fermi-Dirac integrals of order $j = 1/2$ [3]. This has been of particular interest because it can be used to calculate the Fermi level from the known bulk charge density in Eq. (11), as $\eta_F = \mathcal{F}^{-1}{}_{1/2}(n/N_{3D})$. Joyce and Dixon [17] examined a series approach that gives $|\Delta\eta_F| \leq 0.01$ for $\eta_{Fmax} \simeq 5.5$ [3], and a simpler expression from Joyce [18] gives $|\Delta\eta_F| \leq 0.03$ for $\eta_{Fmax} \simeq 5$ [3]. Nilsson proposed two different full-range ($-10 \leq \eta_F \leq 20$) expressions [19] with $|\Delta\eta_F| \leq 0.01$ and $|\Delta\eta_F| \leq 0.005$ [3]. Nilsson later presented two empirical approximations [20] that give $|\Delta\eta_F| \leq 0.01$ for $\eta_{Fmax} \simeq 5.5$ and $\eta_{Fmax} \simeq 20$, respectively [3].

## 6. Table-Based Calculation of Fermi-Dirac Integrals and Inverses

See the Preface to the 4th Edition and [21].

## 7. Numerical Evaluation of Fermi-Dirac Integrals

Fermi-Dirac integrals can be evaluated accurately by numerical integration. Here we briefly review the approach by Press *et al*. for generalized Fermi-Dirac integrals with order $j > -1$ [22]. In this approach, the composite trapezoidal rule with variable transformation $\varepsilon = \exp(t - e^{-t})$ is used for $\eta_F \le 15$ , and the double exponential (DE) rule is used for larger $\eta_F$. Double precision (eps('double') = 2^(-52) $\sim 2.2\times 10^{-16}$) can be achieved after 60 to 500 iterations [22]. The Matlab function, "FD_int_num.m" [1], evaluates the Fermi-Dirac integral numerically using the composite trapezoidal rule following the approach in [22]. The source code is listed in the Appendix. This approach provides very high accuracy, but the CPU time is considerably longer. An online simulation tool that calculates the Fermi-Dirac integrals using this source code has been deployed at nanoHUB.org [23]. Note that the numerical approach we consider in this note is relatively simple, and there are other advanced numerical integration algorithms [24] suggested to improve the calculation speed.

In Figs. 1-2, we compare the accuracy and the timing of the four approaches that calculate $\mathcal{F}_j(\eta_F)$. Fermi-Dirac integrals of order $j = -1/2$ ($\mathcal{F}_{-1/2}(\eta_F)$, Fig. 1) and order $j = 1/2$ ($\mathcal{F}_{1/2}(\eta_F)$, Fig. 2) are calculated for $-10 \le \eta_F \le 10$ with $\eta_F$ spacing = 0.01 using approximate expressions ("FD_int_approx.m" and "FDjx.m" [1]), table-based interpolation ("TBFD_integral.m" [21]), and the rigorous numerical integration ("FD_int_num.m" [1]) with double precision. (Note that it is possible to increase the error tolerance to reduce the CPU time for the numerical integration.) The relative errors of the approximate methods are calculated as $\left(\mathcal{F}_{j,approx} - \mathcal{F}_{j,num}\right) / \mathcal{F}_{j,num}$, where $\mathcal{F}_{j,approx}$ and $\mathcal{F}_{j,num}$ represent the results from the approximate method and the numerical integration, respectively. The elapsed time measured for each approach (using Matlab commands "tic/toc" at Intel Core i5 with 8.0 GB RAM) shows the compromise between the accuracy and the CPU time. The new table-based scheme [21] is nearly as fast as the other approximate routines but less likely to produce large errors in a wide range of $\eta_F$. (Matlab scripts to generate the benchmark plots in Figs. 1-2 are also available at [1].)

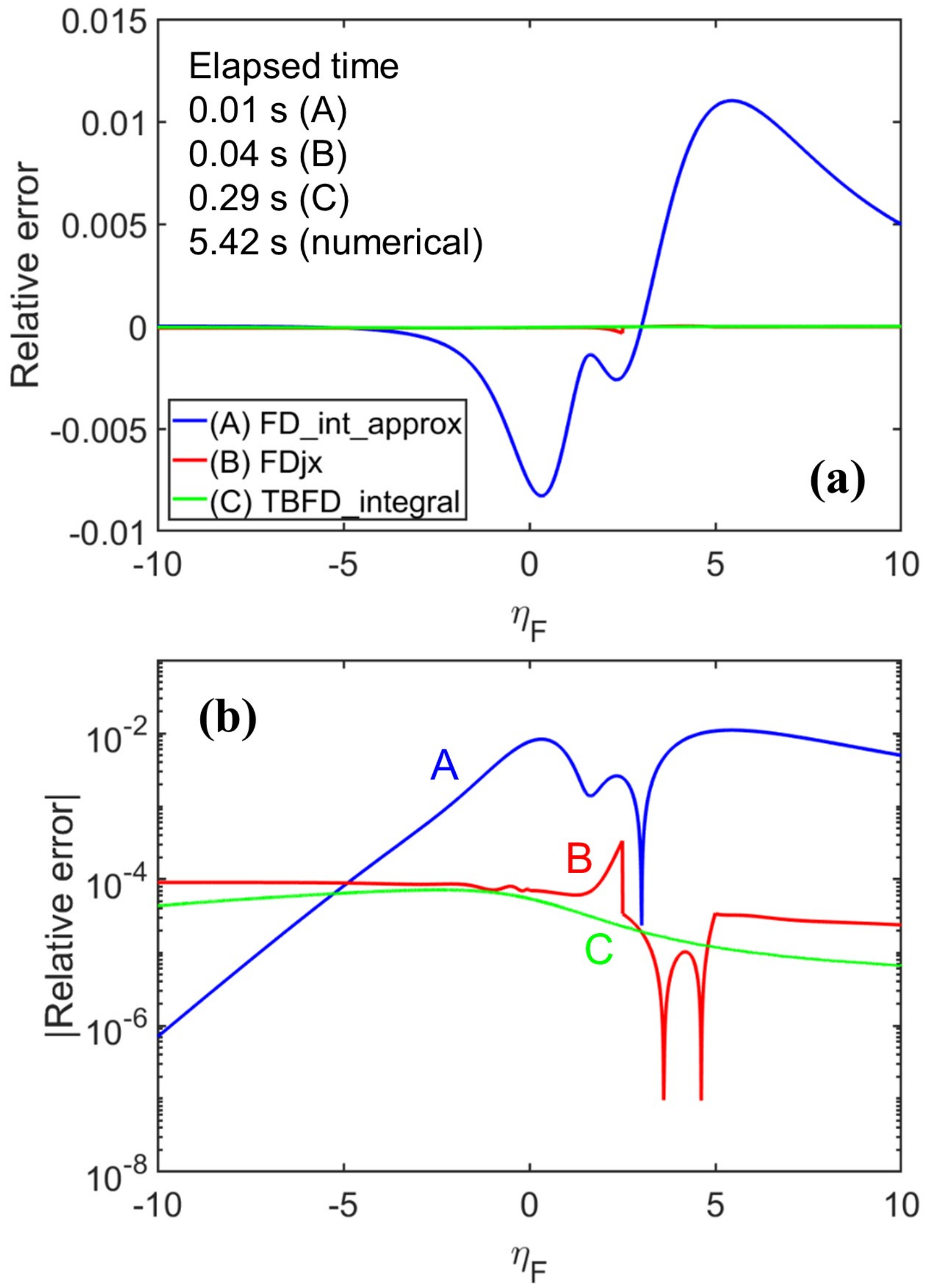

Fig. 1. (a) Relative errors from the approximate expressions for $\mathscr{F}_{-1/2}(\eta_F)$ with respect to the numerical integration ("FD_int_num.m" [1]) with double precision. (A) Relative error of "FD_int_approx.m" [1]. (B) Relative error of "FDjx.m" [1]. (C) Relative error of "TBFD_integral.m" [21]. Matlab scripts to generate plots above are also available at [1]. (b) The absolute values of the relative errors in the log scale. The elapsed time measured for the four approaches shows the trade-off between the accuracy and the CPU time. The new table-based scheme (C) is less likely produce large errors in a wide range of $\eta_F$ while the computation time still remains short.

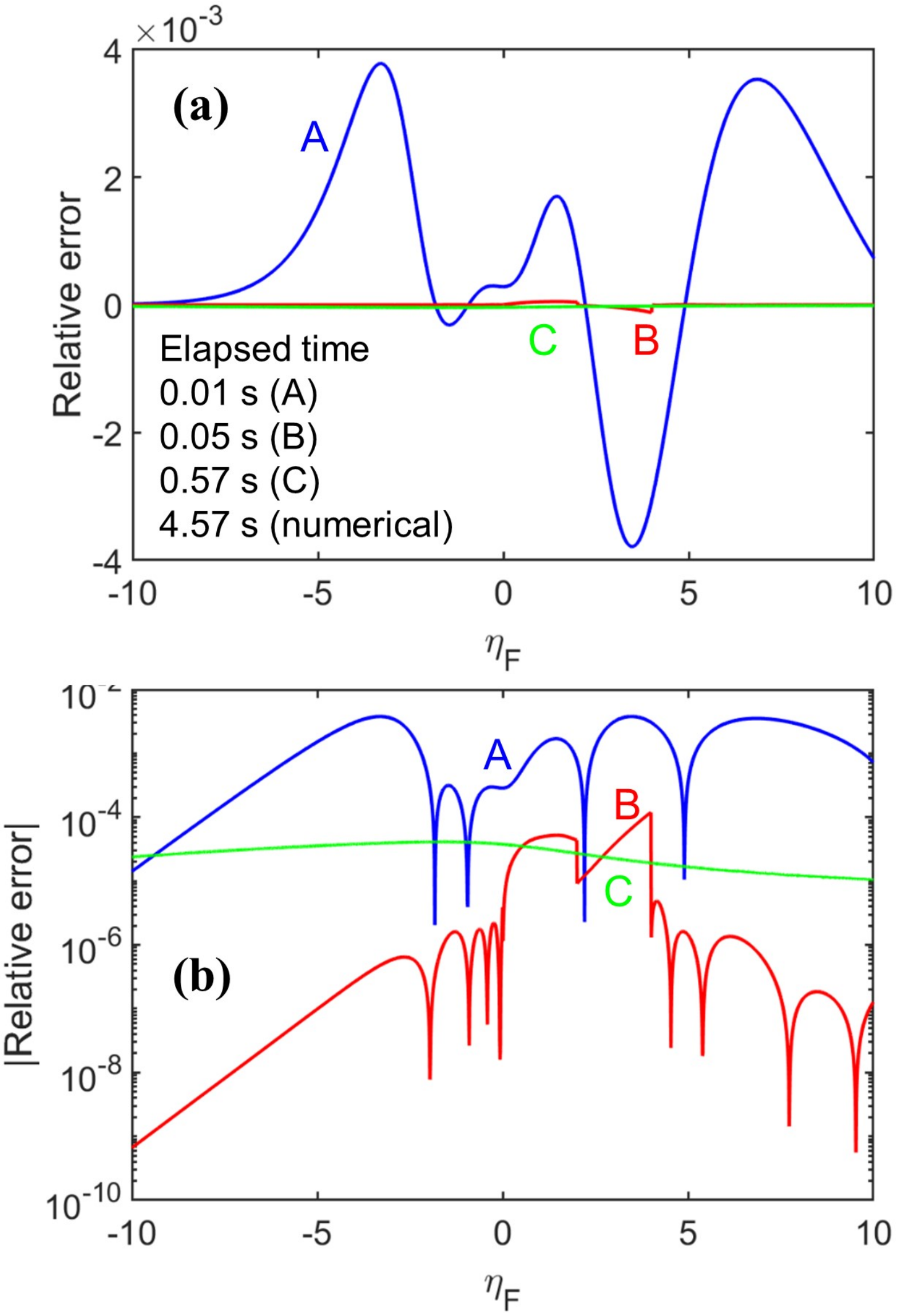

Fig. 2. (a) Relative errors from the approximate expressions for $\mathscr{F}_{1/2}(\eta_F)$ with respect to the numerical integration ("FD_int_num.m" [1]) with double precision. (A) Relative error of "FD_int_approx.m" [1]. (B) Relative error of "FDjx.m" [1]. (C) Relative error of "TBFD_integral.m" [21]. Matlab scripts to generate plots above are also available at [1]. (b) The absolute values of the relative errors in the log scale. The elapsed time measured for the four approaches shows the trade-off between the accuracy and the CPU time. The new table-based scheme (C) is less likely produce large errors in a wide range of $\eta_F$ while the computation time still remains short.

## Appendix

Matlab scripts for the four different approaches to calculate Fermi-Dirac integrals are available at [1] ("FD_int_approx.m", "FDjx.m", "FD_int_num.m") and [21] ("TBFD_integral.m", the latest script generally recommended). Matlab scripts that are relatively short ("FD_int_approx.m" and "FD_int_num.m") are also copied below for readers' information.

"FD_int_approx.m"

```
function y = FD_int_approx( eta, j )

% Analytic approximations for Fermi-Dirac integrals of order j > -1/2
% Date: September 29, 2008
% Author: Raseong Kim (Purdue University)
%
% Inputs
% eta: eta_F
% j: FD integral order
%
% Outputs
% y: value of FD integral
%
% References
% [1]D. Bednarczyk and J. Bednarczyk, Phys. Lett. A, 64, 409 (1978)
% [2]J. S. Blakemore, Solid-St. Electron, 25, 1067 (1982)
% [3]X. Aymerich-Humet, F. Serra-Mestres, and J. Millan, Solid-St. Electron, 24, 981 (1981)
% [4]X. Aymerich-Humet, F. Serra-Mestres, and J. Millan, J. Appl. Phys., 54, 2850 (1983)

if j < -1/2
  error( 'The order should be equal to or larger than -1/2.')
else
  x = eta;
  switch j
    case 0
      y = log( 1 + exp( x ) );        % analytic expression

    case 1/2
      % Model proposed in [1]
      % Expressions from eqs. (22)-(24) of [2]
      mu = x .^ 4 + 50 + 33.6 * x .* ( 1 - 0.68 * exp( -0.17 * ( x + 1 ) .^ 2 ) );
      xi = 3 * sqrt( pi ) ./ ( 4 * mu .^ ( 3 / 8 ) );
      y = ( exp( - x ) + xi ) .^ -1;

    case 3/2
      % Model proposed in [3]
      % Expressions from eq. (5) of [3]
      % The integral is divided by gamma( j + 1 ) to make it consistent with [1] and [2].
      a = 14.9;
      b = 2.64;
      c = 9 / 4;
      y = ( ( j + 1 ) * 2 ^ ( j + 1 ) ./ ( b + x + ( abs( x - b ) .^ c + a ) .^ ( 1 / c ) ) .^ ( j + 1 ) ...
      + exp( -x ) ./ gamma( j + 1 ) ) .^ -1 ./ gamma( j + 1 );

    otherwise
      % Model proposed in [4]
      % Expressions from eqs. (6)-(7) of [4]
      % The integral is divided by gamma( j + 1 ) to make it consistent with [1] and [2].
      a = ( 1 + 15 / 4 * ( j + 1 ) + 1 / 40 * ( j + 1 ) ^ 2 ) ^ ( 1 / 2 );
```

```
        b = 1.8 + 0.61 * j;
        c = 2 + ( 2 - sqrt( 2 ) ) * 2 ^ ( - j );
        y = ( ( j + 1 ) * 2 ^ ( j + 1 ) ./ ( b + x + ( abs( x - b ) .^ c + a ^ c ) .^ ( 1 / c ) ) .^ ( j + 1 ) ...
            + exp( -x ) ./ gamma( j + 1 ) ) .^ -1 ./ gamma( j + 1 );
    end
end
```

## "FD_int_num.m"

```
function [ y N err ] = FD_int_num( eta, j, tol, Nmax )

% Numerical integration of Fermi-Dirac integrals for order j > -1.
% Author: Raseong Kim
% Date: September 29, 208
% Extended (composite) trapezoidal quadrature rule with variable
% transformation, x = exp( t - exp( t ) )
% Valid for eta ~< 15 with precision ~eps with 60~500 evaluations.
%
% Inputs
% eta: eta_F
% j: FD integral order
% tol: tolerance
% Nmax: number of iterations limit
%
% Note: When "eta" is an array, this function should be executed
% repeatedly for each component.
%
% Outputs
% y: value of FD integral
% N: number of iterations
% err: error
%
% Reference
% [1] W. H. Press, S. A. Teukolsky, W. T. Vetterling, and B. P. Flannery,
% Numerical recipies: The art of scientific computing, 3rd Ed., Cambridge
% University Press, 2007.

for N = 1 : Nmax
    a = -4.5;                   % limits for t
    b = 5.0;
    t = linspace( a, b, N + 1 );    % generate intervals
    x = exp( t - exp( -t ) );
    f = x .* ( 1 + exp( -t ) ) .* x .^ j ./ ( 1+ exp( x - eta ) );
    y = trapz( t, f );

    if N > 1                    % test for convergence
        err = abs( y - y_old );
        if err < tol
            break;
        end
    end

    y_old = y;
end

if N == Nmax
    error( 'Increase the maximum number of iterations.')
end

y = y ./ gamma( j + 1 );
```